\documentclass[preprint, aps, showpacs, 12pt]{revtex4}
\usepackage{graphicx}
\usepackage[T1]{fontenc}

\begin{document}
\title{Electron-hole coexistence in disordered graphene probed by high-field magneto-transport}
\author{J.M. Poumirol$^{1}$, W. Escoffier$^{1}$, A. Kumar$^{1}$, M. Goiran$^{1}$, B. Raquet$^{1}$ and J.M. Broto$^{1}$}
\affiliation{$^1$ Laboratoire National des Champs Magnétiques Intenses, University of Toulouse, UPS, INSA, CNRS-UPR 3228, 143 av. de Rangueil, F-31400 Toulouse, France}
\begin{abstract}
We report on magneto-transport measurement in disordered graphene under pulsed magnetic field of up to 57T. For large electron or hole doping, the system displays the expected anomalous Integer Quantum Hall Effect (IQHE) specific to graphene up to filling factor $\nu=2$. In the close vicinity of the charge neutrality point, the system breaks up into co-existing puddles of holes and electrons, leading to a vanishing Hall and finite longitudinal resistance with no hint of divergence at very high magnetic field. Large resistance fluctuations are observed near the Dirac point. They are interpreted as the the natural consequence of the presence of electron and hole puddles. The magnetic field at which the amplitude of the fluctuations are the largest is directly linked to the mean size of the puddles.
\end{abstract}

\maketitle

A few years ago, a new form of IQHE has been discovered in graphene for which the Hall conductance is half-integer quantized in units of $4.e^2/h$ \cite{Novoselov2005}. The factor $4$ stands for Landau level (L.L.) degeneracy (two-fold spin and two-fold nodal degeneracy) so that IQHE occurs at filling factor $\nu=4\times\left(|i|+1/2\right)$, where $i=0, \pm1, \pm2 ...$ is an integer used for L.L. indexation. In very high magnetic field and close to the Dirac point, spin and nodal degeneracies are expected to be lifted. Indeed, pioneering experimental works \cite{Zhang2006, Jiang2007} showed additional Hall resistance plateaus for filling factor $\nu=0$, $\nu=\pm1$, $\nu=\pm4$ as well as $\pm3$. While Zeeman coupling lifts spin degeneracy, the nodal degeneracy lifting implies spontaneous symmetry breaking effects \cite{Yang2007, Alicea2007} which fundamental nature is still under investigation \cite{Alicea2006, Nomura2006, Abanin2007, Gusynin2008, Gorbar2008, Giesbers2009, Zhang2009}. It is now established that disorder significantly affects the electronic properties of graphene and thus cannot be ignored, especially at low charge carrier density where it brings about the formation of electron-hole puddles \cite{Martin2008}. Even if the field-induced insulator transition is well documented for clean graphene \cite{Jung2009, Amado2009}, little is known about high-field transport in disordered samples where the presence of electron-hole puddle is expected to dominate over a large magnetic field range \cite{Sarma2009}.\\

In this communication, we report on IQHE in a disordered system for various charge carrier densities from hole to electron doping. At low field, our experimental data are consistent with existing theories for disordered graphene. Close the the Dirac point, the magneto-resistance is quasi-flat for a broad magnetic field range without sign of divergence in contrast with the clean case. In addition, at large magnetic field, we show that both longitudinal and Hall magneto-resistance display large fluctuations at the Dirac point, whith a maximum amplitude for $B\approx 25 T$. We show that these fluctuations result from the presence of electron-hole puddles with an estimated mean diameter of $5 nm$.


\begin{figure}[!ht]
\includegraphics[width=12cm]{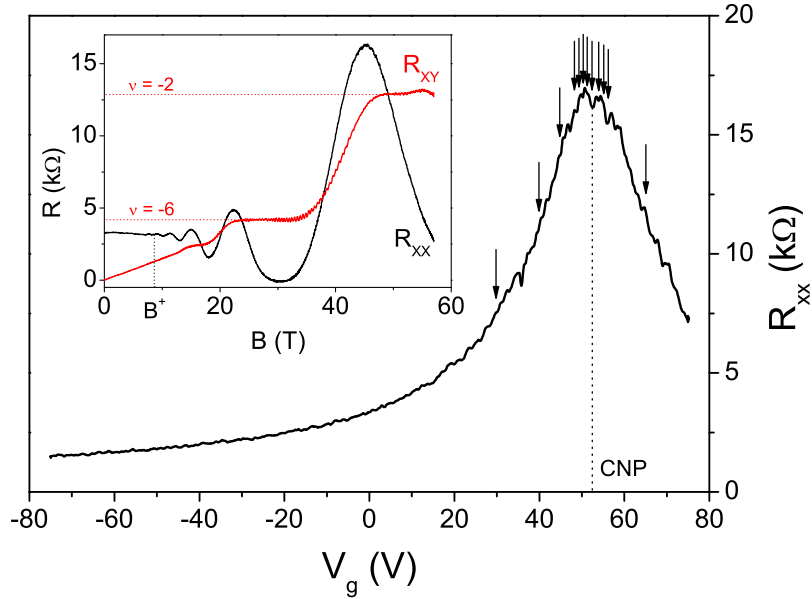}
\caption{\label{fig.1} Longitudinal resistance as a function of gate voltage $V_g$ at $T=1.5K$ and $B=0T$, the charge neutrality point is located at $V_g=+52V$. The arrows indicate specific values of gate voltage at which magneto-resistance data will be presented (see figure \ref{fig.3}). Inset : Longitudinal resistance $R_{xx}$ and Hall resistance $R_{xy}$ as a function of magnetic field at $T=1.5K$ and $V_g=0V$.}
\end{figure}

The sample has been designed in the standard Hall bar geometry using electron-beam lithography, allowing simultaneous measurements of both Hall $(R_{xy})$ and longitudinal $(R_{xx})$ resistances. Variable electrostatic doping is achieved upon applying a voltage through $300 nm$ of $SiO_2$ dielectric used as a back-gate. Figure \ref{fig.1} shows the device resistance as the gate voltage $V_g$ is ramped from $-70V$ to $+70V$. The charge neutrality point is identified at the resistance maximum $V_{CNP}=+52V$. The large FWHM of the curve as well as the large shift of $V_{CNP}$ from $V_g=0V$ suggests a rather disordered sample. Actually, non-intentional hole doping certainly arises from defects and charged chemical species adsorbed at the surface of the sample or at the $Si0_2$ interface during its fabrication or exposure to air. The gating efficiency of the device is estimated to $\beta=7.85\times 10^{10}cm^{-2}/V$ from low field Hall measurements. This result is compared to the accepted plane capacitor model for graphene $\beta=(\epsilon_0\epsilon_r)/(e.d)$ where $\epsilon_0\epsilon_r$ is the dielectric permittivity of $SiO_2$ and $d=275 nm$ its thickness, in good agreement with the nominal value. A constant current of $i=0.5\mu A$ is injected through the device at $1.5K$ while a strong pulsed magnetic field, of total duration $300ms$ and rising up to $57T$, is applied perpendicular to the sample. Inset of figure \ref{fig.1} shows IQHE at $V_g=0V$. The Hall resistance displays well defined quantized plateaus at $R_{xy}=\frac{h}{4e^2}.(i+1/2)^{-1}$ for $i=0,1,2...$ typical for graphene. The mean mobility is estimated to $\mu=1300 cm^2/V.s$ in consistency with the first appearance of SdH oscillations at minimum magnetic field $B^\dagger\approx7T$ (see inset of figure \ref{fig.1}).\\

\begin{figure}[!ht]
\includegraphics[width=12cm]{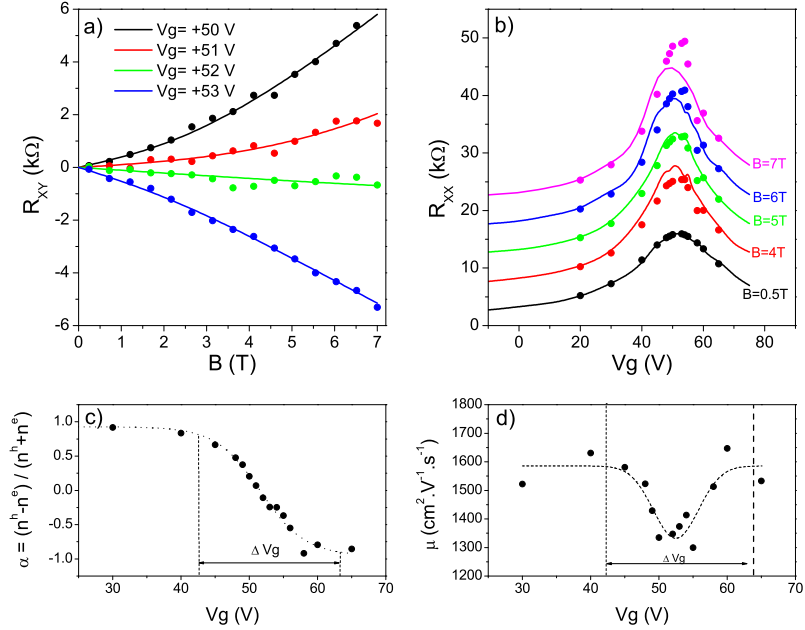}
\caption{\label{fig.2} (a) and (b) Theoretical adjustment of $R_{xy}(B)$ and $R_{xx}(V_g)$ at small field respectively, at selected back gate voltages or magnetic field using the two-fluid model. $R_{xx}(V_g)$ data have been successively upshifted of $\Delta R=5k\Omega$ for clarity. Note that at $B=B\dagger=7T$, the two-fluid model is no longer accurate (c) and (d) Extracted coefficient $\alpha$ and mobility $\mu$ as a function of $V_g$ : notice the coexistence of electrons and hole-like particles for a broad range of gate voltage $\Delta V_g\approx 23V$, as well as a mobility dip near CNP. Dotted lines are guide for the eye.}
\end{figure}

For $B<B^{\dagger}$, in the classical regime, the experimental data can be approached using a standard two-fluid model \cite{Hwang2007-2, Rossi2009}. In this picture, electron and hole-like particles coexist and both contribute to transport. For simplicity and within a good approximation, we assume the same mobility $\mu$ for both electrons and holes. Their density will be denoted $n_e$ and $n_h$ respectively. In the following, we use the reduced parameter $\alpha=(n_h-n_e)/(n_e+n_h)$ so that $R_{xx}(B)=R_{xx}(0)\times\frac{1+\left(\mu.B\right)^2}{1+\left(\alpha.\mu.B\right)^2}$ and $R_{xy}=\frac{1}{\gamma}\alpha.\mu.B.R_{xx}(B)$ with $R_{xx}(0)=\gamma\times\left[(n_e+n_p).e.\mu\right]^{-1}$ and $\gamma=3$ is the form factor \cite{Cho2008}. A theoretical adjustment with $\alpha$ and $\mu$ as free parameters is performed to match the full experimental dataset. The agreement is convincing as shown in figure \ref{fig.2}-a and \ref{fig.2}-b. We emphasis on the fact that $R_{xx}(0)$ is not a free parameter and is determined from the $R_{xx}(V_g)$ curve at zero magnetic field (see figure \ref{fig.1}). Although some refinements (e.g. different mobilities for electrons and holes particles) would certainly improve the fitting quality, both the longitudinal and Hall resistance can be adjusted with this minimal two-fluid model which captures well the underlying physics of disordered graphene. This fitting procedure allows us to extract the profiles of $\alpha(V_g)$ and $\mu(V_g)$ (see figures \ref{fig.2}-c and \ref{fig.2}-d). On crossing the charge neutrality point, the hole to electron density ratio changes smoothly within a gate voltage interval $\Delta V_g\approx 20V$. In the presence of disorder, mainly modeled by charged impurities, the coexistence of hole and electron-like particles is expected within a window $\Delta V_g=\frac{2.n^*}{\beta}$ where $n^*$ is an effective charge carrier density defined by $n^*=\frac{\left(5\times10^{15}/\mu\right)^2}{4.\beta.V_{CNP}}$ \cite{Hwang2007}. We find $\Delta V_g\approx 23V$ in good agreement with experimental results. The predicted minimum of conductivity at CNP and zero field is $G_{CNP}=\frac{1}{\gamma}\times\frac{20.e^2}{h}\times\frac{n^*.\mu}{5.10^{15}}=61 \mu S$ which is favorably compared with experiment (i.e. $G^{exp}_{CNP}=60.6 \mu S$). In first approximation, graphene's resistivity was experimentally found to be inversely proportional to the density of charge carriers, so that their mobility is almost independent of $n\propto V_g$. This particularly holds for low mobility samples, far from the Dirac point, where conduction is mainly dominated by long-range scatterers such as charged impurities. On the other hand, little is known about the carrier's mobility near the CNP since, when calculated using simple Drude model, $\mu=\sigma/e.n$ diverges for $n\rightarrow0$. Using the two-fluid model fitting procedure, the extracted curve $\mu(V_g)$ in figure \ref{fig.2}-d shows a clear minimum of the mobility in the vicinity of the CNP. Do and Dollfus \cite{Do2009} compared the effects of long-range and short-range scattering centers on mobility. They demonstrated that, even if the mobility is peaked at $V_g=V_{CNP}$ in the presence of charged impurities, it is strongly reduced when vacancies are included, leading to a dip-like curve consistent with our observations.\\

\begin{figure}[!ht]
\includegraphics[width=12cm]{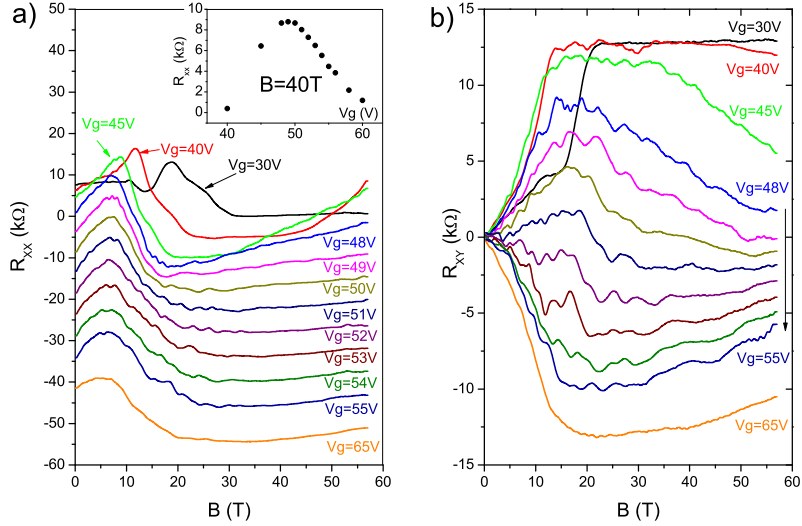}
\caption{\label{fig.3} Longitudinal resistance (a) and Hall resistance (b) as a function of magnetic field for selected gate voltages (see arrows in figure \ref{fig.1}) in the vicinity of the charge neutrality point. Note that in (a), the curves have been successively down-shifted of $-5 k\Omega$ with respect to curve at $V_g=30V$ for clarity. Inset shows $R_{xx}(B=40T)$.}
\end{figure}

We now turn our attention to high magnetic field transport properties. Figure \ref{fig.3}-a and \ref{fig.3}-b show the longitudinal and Hall resistances in the vicinity of the charge neutrality point. Focusing close to $V_{CNP}\approx 52V$, the longitudinal magneto-resistance is first positive and reaches a maximum for $B\approx B^{\dagger}=7T$. This resistance maximum is attributed to the formation of overlapping LLs and naturally sets the crossover from the classical to the quantum regime. The resistance then decreases as a consequence of increasing LLs energy interval with magnetic field till the $i=0$ LL only remains populated. From $B\approx 20T$, a merely flat magneto-resistance is observed up to the highest experimentally available magnetic field. In graphene, the energy of the $n=0$ LL is independent of magnetic field. Therefore, as long as long as the Fermi energy remains at a constant value within the $n=0$ Landau Level, a fixed back-scattering rate between edge states through extended states accounts for the quasi-flat magneto-resistance. The scattering rate is maximum when $V_g=V_{CNP}$, that is when the Fermi energy is aligned with the center of the $n=0$ L.L. (see inset of figure \ref{fig.3}). As a matter of fact, instead of a purely flat magneto-resistance, a weak increase is observed at very high magnetic field. The closer $V_g$ is to $V_{CNP}$, the less pronounced is the high field magneto-resistance increase. This effect is explained considering a Fermi energy shift towards the center of the $n=0$ LL as its degeneracy increases with the field. The energy shift of $E_F$ is all the more important as the initial density of states is weak. We note that, contrary to clean graphene devices \cite{Checkelsky2009}, no divergence of the magneto-resistance is observed up to $B=57T$ for this particular disordered device. This figure, together with the above-mentioned $R_{xx}(B,V_g)$ behavior, rule out any alternative explanations implying spontaneous symmetry breaking and degeneracy lifting. However, whether a divergent behavior will be detected or not for even higher magnetic field remains an open question. In the close vicinity of the CNP, the Hall resistance displays no more well defined quantized plateaus (e.g. compared with $R_{xy}(B)$ curves at $V_g=30V$ showing well developed Hall plateau at $R_{xy}=12.9k\Omega$), but remains close to zero as a result of electron and hole balance. We note that, for some fixed values of the gate voltage, the Hall resistance experiences a sign inversion indicating a change of the majority carrier type. This effect is related to a field-enhanced coexisting energy zone of electron and hole particles and will be discussed elsewhere.\\

\begin{figure}[!ht]
\includegraphics[width=12cm]{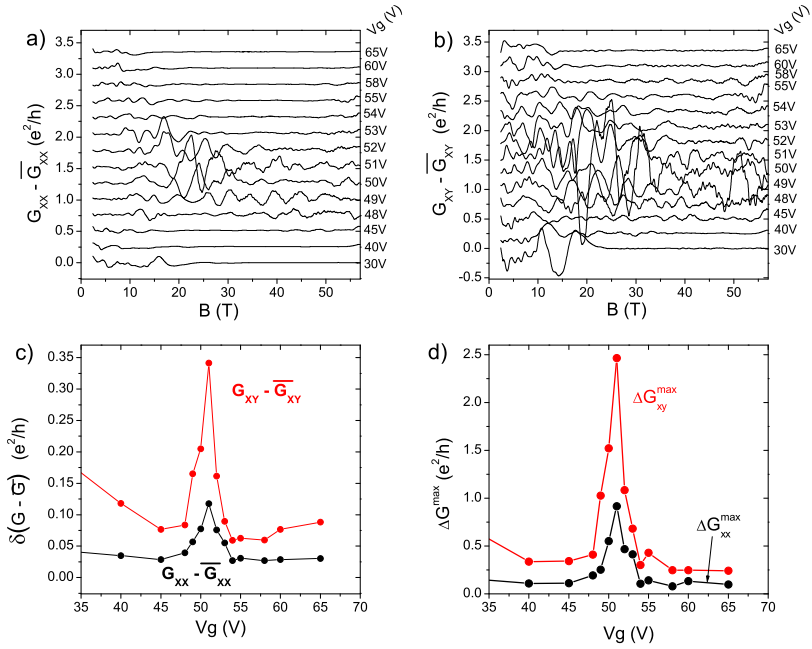}
\caption{\label{fig.4} (a) and (b) : Longitudinal and Hall resistance fluctuations for various gate voltages. A smooth background function has been subtracted to the original data. Data have been shifted for clarity. (c) Standard deviation of the curves shown in a) (black) and b) (red). d) Maximum peak-to-peak amplitude of the fluctuations.}
\end{figure}

In the vicinity of the Dirac point, the Hall and longitudinal magneto-resistances are distorded with large fluctuations. Such fluctuating features are qualitatively analyzed in the conductance data, obtained by resistance tensor inversion. Figures \ref{fig.4}-a) and \ref{fig.4}-b) highlight on such conductance fluctuations, where a smooth background function $\overline{G}(B)$ has been subtracted to the original conductance data. The standard deviation as a function of gate voltage has been displayed in figure \ref{fig.4}-c). The amplitude of the fluctuations are much larger in the vicinity of the CNP (figure \ref{fig.4}-c) and in the intermediate magnetic field range ($10T<B<35T$). Similar fluctuations have already been reported in various DC high magnetic-field experimental works \cite{Zhang2006, Abanin2007, Giesbers2009} as a function of gate voltage, but they are not systematically observed \cite{Checkelsky2008} depending on sample quality, magnetic field range and temperature. In disordered materials, magneto-resistance fluctuations are usually understood in terms of magnetic-field dependent phase shifts of the electronic wave-functions and are well reproduced using the UCF theory \cite{Morozov2006, Tikhonenko2008}. However, their occurrence in the vicinity of the Dirac point only and their large amplitude up to $2.5 e^2/h$ call for an alternative explanation. In disordered graphene, close to the Dirac point, the mean carrier density is low and screening is reduced so that the surrounding disordered medium drives the system into a fluctuating potential landscape. This leads to the natural formation of co-existing local electron and hole puddles \cite{Martin2008}. Following the lines of \cite{Hwang2007, Adam2007}, disorder can be generally described through a charged impurity density $n_{imp}$ that scales as $n_{imp}=5\times10^{15}/\mu$. Using mobility $\mu=1300 cm^2/V.s$ for our sample, we compute $n_{imp}=3.85\times 10^{12}cm^{-2}$, so that the typical mean scattering length is roughly $5nm$. Using a scanning single-electron transistor, Martin {\it et. al.} \cite{Martin2008} have demonstrated that the mean puddle dimensions are of the same order as the typical disorder length. It is therefore reasonable to consider a size distribution of electron-hole regions centered at $\ell_{p}\approx 5\ nm$. When a magnetic field is applied to such an inhomogeneous system, within a specific puddle charge carriers tend to be quantized in cyclotron orbits provided the puddle size is larger than the magnetic length $\ell_B=\sqrt{\frac{\hbar}{e.B}}$. Therefore, as the magnetic field increases, the puddles successively enter into the quantum regime depending on their sizes. This process leads to conductance fluctuations of maximum amplitude expected for $\ell_B=\ell_p$ i.e. $B=25$T as shown in figures \ref{fig.4}-a) and b). This result should be compared with the work of Branchaud {\it et. al.} \cite{Branchaud2010} who analysed magneto-conductance fluctuations close to CNP and deduced a mean puddle size of about $125 nm$ for clean graphene. Also, Rycerz {\it et. al.} \cite{Rycerz2007} interpreted the conductance fluctuations as a natural consequence of the presence of electron-hole puddles, invoking a percolation regime in which the magnetic field induces small variations of the charge carrier trajectories.

To conclude, we have investigated the electronic properties of disordered graphene in the close vicinity of the Dirac point. At low magnetic field, the system is well described using a simple two fluid model where both electron and hole-like particles contribute to transport. The co-existence energy interval is well accounted by existing theory for disordered graphene whereas the mobility dip at low carrier density is possibly explained by the presence of vacancies. When the magnetic field is high enough so that the energy spectrum displays a clear LL quantization, large fluctuations of the magneto-resistance near the Dirac point are observed and explained in terms of topological transition of electron-hole puddles into the quantum regime. These fluctuations progressively vanishes as the magnetic field is further increased, suggesting a conduction regime where the magnetic field length is smaller than the mean puddle size. For clean graphene, spin and valley degeneracy splitting occur in strong magnetic field and, for the cleanest samples, the fractional quantum Hall effect \cite{Du2009, Bolotin2009} has even been observed. Many experimental studies converge to the assumption that the more disordered graphene is, the highest is the magnetic field at which such effects occur. It makes no doubt that similar experiments with samples having different degrees of disorder will significantly enlarge our understanding of the dirty regime.\\

We are embedded to K.S. Novoselov for providing sample. This work was funded by the french National Agency for Research (ANR) through the program referenced ANR-08-JCJC-0034-01 and EUROMAGNET under E.U. contract n$^o$ 228043.\\

\bibliographystyle{unsrt}
\bibliography{escoffier1304}{1}

\begin{thebibliography}{10}

\bibitem{Novoselov2005}
K.S. Novoselov, A.K. Geim, S.V. Morozov, D.~Jiang, M.I. Katsnelson, I.V.
  Grigorieva, S.V. Dubonos, and A.A. Firsov.
\newblock Two-dimensional gas of massless dirac fermions in graphene.
\newblock {\em Nature}, 438:197, 2005.

\bibitem{Zhang2006}
Y.~Zhang, Z.~Jiang, J.P. Small, M.S. Purewal, Y.-W. Tan, M.~Fazlollahi, J.D.
  Chudow, J.A. Jaszcak, H.L. Stormer, and P.~Kim.
\newblock Landau level splitting in graphene in high magnetic field.
\newblock {\em Phys. Rev. Lett.}, 96:136806, 2006.

\bibitem{Jiang2007}
Z.~Jiang, Y.~Zhang, H.L. Stormer, and P.~Kim.
\newblock Quantum hall states near the charge-neutral dirac point in graphene.
\newblock {\em Phys. Rev. Lett.}, 99:106802, 2007.

\bibitem{Yang2007}
K.~Yang.
\newblock Spontaneous symmetry breaking and quantum hall effect in graphene.
\newblock {\em Solid State Commun.}, 143:27, 2007.

\bibitem{Alicea2007}
J.~Alicea and M.~P.A. Fisher.
\newblock Interplay between lattice-scale physics and the quantum hall effect
  in graphene.
\newblock {\em Solid State Commun.}, 143:504, 2007.

\bibitem{Alicea2006}
J.~Alicea and M.P.A. Fisher.
\newblock Graphene integer quantum hall effect in the ferromagnetic and
  paramagnetic regimes.
\newblock {\em Phys. Rev. B}, 74:075422, 2006.

\bibitem{Nomura2006}
K.~Nomura and A.H.~Mc Donald.
\newblock Quantum hall ferromagnetism in graphene.
\newblock {\em Phys. Rev. Lett.}, 96:256602, 2006.

\bibitem{Abanin2007}
D.~A. Abanin, K.S. Novoselov, U.~Zeitler, P.A. Lee, A.K. Geim, and L.S.
  Levitov.
\newblock Dissipative quantum hall effect in graphene near the dirac point.
\newblock {\em Phys. Rev. Lett.}, 98:196806, 2007.

\bibitem{Gusynin2008}
V.P. Gusynin, V.A. Miransky, S.G. Sharapov, and I.A. Shovkovy.
\newblock Edge states, mass and spin gaps, and quantum hall effect in graphene.
\newblock {\em Phys. Rev. B}, 77:205409, 2008.

\bibitem{Gorbar2008}
E.V. Gorbar and V.P. Gusynin.
\newblock Towards a theory of the quantum hall effect in graphene.
\newblock {\em Low Temp. Phys.}, 34:790, 2008.

\bibitem{Giesbers2009}
A.J.M. Giesbers, L.A. Ponomarenko, K.S. Novoselov, A.K. Geim, M.I. Katsnelson,
  J.C. Maan, and U.~Zeitler.
\newblock Gap opening in the zeroth landau level of graphene.
\newblock {\em Phys. Rev. B}, 80:201403(R), 2009.

\bibitem{Zhang2009}
Y.Y. Zhang, J.~Hu, B.A. Bernevig, X.R. Wang, X.C. Xie, and W.M. Liu.
\newblock Localization and the kosterlitz-thouless transition in disordered
  graphene.
\newblock {\em Phys. Rev. Lett.}, 102:106401, 2009.

\bibitem{Martin2008}
J.~Martin, N.~Akerman, G.~Ulbricht, T.~Lohmann, J.H. Smet, K.~von Klitzing, and
  A.~Yacoby.
\newblock Observation of electron-hole puddles in graphene using a scanning
  single-electron transistor.
\newblock {\em Nature Physics}, 4:144, 2008.

\bibitem{Jung2009}
J.~Jung and A.H. MacDonald.
\newblock Theory of the magnetic-field-induced insulator in neutral graphene
  sheets.
\newblock {\em Phys. Rev. B}, 80:235417, 2009.

\bibitem{Amado2009}
M.~Amado, E.~Diez, D.~Lopez-Romero, F.~Rossella, J.M. Caridad, V.~Bellani, and
  D.K. Maude.
\newblock Universal scaling of the metal-insulator transition for filling
  factor nu=2 to nu=0 in graphene.
\newblock {\em arXiv:0907.1492v1 [cond-mat.mes-hall]}, 2009.

\bibitem{Sarma2009}
S.~Das Sarma and K.~Yang.
\newblock The enigma of $\nu=0$ quantum hall effect in graphene.
\newblock {\em Solid State Commun.}, 149:1502, 2009.

\bibitem{Hwang2007-2}
E.H. Hwang, S.~Adam, and S.~Das Sarma.
\newblock Transport in chemically doped graphene in the presence of adsorbed
  molecules.
\newblock {\em Phys. Rev. B}, 76:195421, 2007.

\bibitem{Rossi2009}
E.~Rossi, S.~Adam, and S.~Das Sarma.
\newblock Effective medium theory for disordered two-dimensional graphene.
\newblock {\em Phys.Rev. B}, 79:245423, 2009.

\bibitem{Cho2008}
S.~Cho and M.S. Fuhrer.
\newblock Charge transport and inhomogeneity near the minimum conductivity
  point in graphene.
\newblock {\em Phys. Rev. B}, 77:081402, 2008.

\bibitem{Hwang2007}
E.H. Hwang, S.~Adam, and S.~Das Sarma.
\newblock Carrier transport in two-dimensional graphene layers.
\newblock {\em Phys. Rev. Lett.}, 98:186806, 2007.

\bibitem{Do2009}
V.~Nam Do and P.~Dollfus.
\newblock Effects of charged impurities and lattice defects on transport
  properties of nanoscale graphene structures.
\newblock {\em J. Appl. Phys.}, 106:023719, 2009.

\bibitem{Checkelsky2009}
J.G. Checkelsky, L.~Liu, and N.P. Ong.
\newblock Divergent resistance at the dirac point in graphene: evidence for a
  transition in a high magnetic field.
\newblock {\em Phys. Rev. B}, 79:115434, 2009.

\bibitem{Checkelsky2008}
J.G. Checkelsky, L.~Liu, and N.P. Ong.
\newblock Zero-energy state in graphene in high magnetic field.
\newblock {\em Phys. Rev. Lett.}, 100:206801, 2008.

\bibitem{Morozov2006}
S.~V. Morozov, K.~S. Novoselov, M.~I. Katsnelson, F.~Schedin, L.~A.
  Ponomarenko, D.~Jiang, and A.~K. Geim.
\newblock Strong suppression of weak localization in graphene.
\newblock {\em Phys. Rev. Lett.}, 97:016801, 2006.

\bibitem{Tikhonenko2008}
F.~V. Tikhonenko, D.~W. Horsell, R.~V. Gorbachev, and A.~K. Savchenko.
\newblock Weak localization in graphene flakes.
\newblock {\em Phys. Rev. Lett.}, 100:056802, 2008.

\bibitem{Adam2007}
S.~Adam, E.H. Huang, V.M. Galitski, and S.~Das Sarma.
\newblock A self consistent theory for graphene transport.
\newblock {\em P.N.A.S.}, 104:18392, 2007.

\bibitem{Branchaud2010}
S.~Branchaud, A.~Kam, P.~Zawadzki, F.~M. Peeters, and A.S. Sachrajda.
\newblock Transport detection of quantum hall fluctuations in graphene.
\newblock {\em Phys. Rev. B}, 81:121406(R), 2010.

\bibitem{Rycerz2007}
A.~Rycerz, J.~Tworzydlo, and C.W.J. Beenakker.
\newblock Anomalously large conductance fluctuations in weakly disordered
  graphene.
\newblock {\em EPL}, 79:57003, 2007.

\bibitem{Du2009}
X.~Du, I.~Skachko, F.~Duerr, A.~Luican, and E.Y. Andrei.
\newblock Fractional quantum hall effect and insulating phase of dirac
  electrons in graphene.
\newblock {\em Fractional quantum Hall effect and insulating phase of Dirac
  electrons in graphene}, 462:192, 2009.

\bibitem{Bolotin2009}
K.I. Bolotin, F.~Ghahari, M.D. Shulman, H.L. Stormer, and P.~Kim.
\newblock Observation of the fractional quantum hall effect in graphene.
\newblock {\em Nature}, 462:196, 2009.

\end{thebibliography}
\end{document}